\documentclass[preprint,5p,10pt]{elsarticle}

\usepackage{lineno}
\usepackage{xcolor}
\usepackage{etoolbox}
\usepackage[normalem]{ulem}
\usepackage{dingbat}
\usepackage{url}

\newcommand{\topowmath}[2][0]{\ifnumcomp{0}{=}{#1}{}{#1~\cdot~}10^{#2}}
\newcommand{\topow}[2][0]{\ifnumcomp{0}{=}{#1}{}{$#1~\cdot$~}$10^{#2}$}

\newcommand{\invmcube}[2][0]{\topow[#1]{#2}~m$^{-3}$}
\newcommand{\invmcubemath}[2][0]{\topowmath[#1]{#2}~\text{m}^{-3}}

\newcommand{\frachalf}{\frac{1}{2}}
\newcommand{\halfdt}{\frac{dt}{2}}
\newcommand{\qmr}[1]{\frac{q_{#1}}{m_{#1}}}
\usepackage{amsmath}
\usepackage{bm}
\usepackage{cancel}
\usepackage{subfig}
\usepackage[T1]{fontenc}

\newcounter{bla}

\begin{document}

\begin{frontmatter}

\title{Particle-In-Cell Simulations of Quantum Plasmas}

\author[a]{Gregory K. Ngirmang\corref{ca}\fnref{firstauth}}
\ead{ngirmang@protonmail.com}
\author[b]{Hue T.B. Do\fnref{firstauth}}
\author[a]{Guangxin Liu}
\author[b]{Michel Bosman}
\author[a]{Lin Wu\corref{ca}}
\ead{lin\_wu@sutd.edu.sg}

\address[a]{Science, Mathematics, and Technology Pillar, Singapore University of Technology and Design, 8 Somapah Road, Singapore 487372
}
\address[b]{Department of Materials Science and Engineering, National University of Singapore, 9 Engineering Drive 1, Singapore 117575
}

\fntext[firstauth]{Gregory K. Ngirmang and Hue T. B. Do are equally the first authors of this work}
\cortext[ca]{Corresponding authors}

\begin{abstract}
Room-temperature metals and semi-metals which consist of a gas of bound electrons in a near-continuum band structure can be classified as cold quantum plasmas. This insight suggests that Particle-in-Cell (PIC) simulations, traditionally used for modeling classical plasmas, may be adapted for the next generation of nanoscopic simulations in photonics, plasmonics, and beyond. This article introduces four key physics modules implemented in two open-source PIC codes that can be applied to condensed matter calculations. These modules include (I) the incorporation of Fermi-Dirac (FD) electrons, (II) material structure boundary conditions, (III) a bound particle model for linear dispersive materials, and (IV) the inclusion of massless Dirac carriers for simulating graphene-like materials. By integrating these modules into existing PIC frameworks, we provide a versatile and self-consistent approach for simulating condensed matter systems, opening new avenues for modeling dynamic phenomena in photonics and plasmonics.
\end{abstract}

\end{frontmatter}

\section{Introduction}

\begin{figure*}[h]
    \centering
        \includegraphics[width=7in]{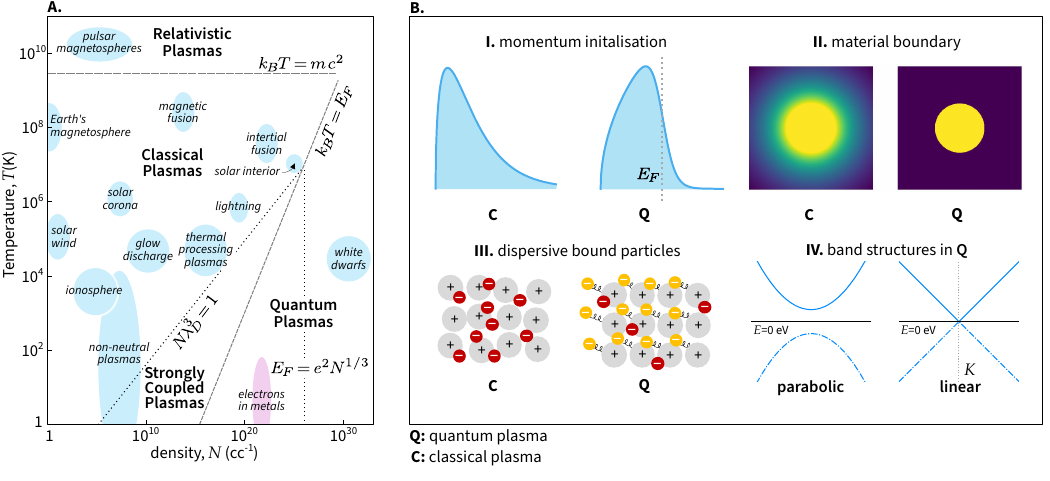}
    \caption{    
    Particle-in-cell (PIC) modeling of quantum plasmas. (A) A diagram illustrating different plasma regimes as a function of temperature and density, with quantum plasmas occupying the high-density, low-temperature region. This corresponds to the domain of condensed, low-temperature electron gases typically observed in solid, low-temperature materials.
    (B) A summary of the key physics modules relevant for simulating quantum plasmas in the condensed matter regime:
    (I) Initialization of charged species with a Dirac-Fermi energy distribution;
    (II) Boundary conditions applied to charged species at the edges of material structures within the simulation space;
    (III) Inclusion of bound species to provide a linear dispersive dielectric response from bound electrons;
    (IV) Modeling of Dirac fermions using macroparticles, enabling simulations of materials with a $K$-point band structure, such as graphene.
    (A) is adapted from \emph{Plasma Science} by the National Research Council, 1995~\cite{nrc1995}}.
    \label{fig:overview}
\end{figure*}

Since the 1940s, computational simulations have dramatically reshaped the landscape of physics, enabling us to explore complex systems beyond the reach of traditional analytical methods. One area where this transformation has been particularly impactful is electromagnetism, which governs the vast majority of practical physical phenomena. At the heart of electromagnetism lies the interaction between charged particles, and, broadly speaking, plasmas—systems of semi-free charged species. While plasmas are often associated with extreme conditions, such as those found in astrophysical phenomena (\textcolor{blue}{Fig.~\ref{fig:overview}A}), many solid materials, including metals and semi-metals, also exhibit quasi-free electrons within a near-continuum of quantum levels. In this context, such materials can be viewed as quantum plasmas, where the charge carriers behave similarly to plasma species, but within a condensed matter setting.

To model complex systems, computational methods such as time-domain and frequency-domain solvers are commonly used. The Finite-Difference Time-Domain (FDTD) method \cite{taflovehagness, teixiera2023} is a widely adopted time-domain approach, while the Finite Element Method (FEM) \cite{liu2022} dominates steady-state frequency-domain simulations. Advanced variants, including potential-based solvers for low-frequency limits \cite{zhang2024}, have further enhanced our understanding of electromagnetic phenomena, from radio waves to quantum plasmonics \cite{esteban2012bridging, wu2013fowler, wu_charge_2016}. These methods generally model material responses as static dielectric functions, which works well for many applications. However, this approach becomes limiting when materials undergo dynamic evolution, affecting electromagnetic interactions such as re-emission and absorption, particularly in quantum plasmas where electron dynamics are non-trivial and significantly impact system behavior.

The concept of quantum plasmas—applied to condensed matter systems—suggests that these systems can be modeled more effectively by considering the quantum nature of their constituent particles. This leads us to the Particle-in-Cell (PIC) method~\cite{birdsalllangdon}, a powerful technique traditionally used to simulate plasmas. In PIC simulations, the coupling between Maxwell's equations (which govern the electromagnetic field) and the motion of charged particles (which follow the Lorentz force law) is explicitly modeled. Instead of using a continuous charge distribution, PIC employs macroparticles—representative particles that capture the behavior of large numbers of real particles. The self-consistent evolution of both the electromagnetic field and particle dynamics makes PIC ideal for simulating quantum plasmas in time-dependent scenarios~\cite{ding_particle_2020, do2021, do2024}.

However, the standard PIC codes have been primarily developed to model classical, weakly coupled plasmas—those typically found in high-temperature or low-density environments. For quantum plasmas, especially in the denser, colder regions typical of condensed matter systems (\textcolor{blue}{Fig. \ref{fig:overview}A}), additional physical models are required. This article focuses on enhancing two popular open-source PIC codes, \textsc{epoch} \cite{Arber2015} and Smilei~\cite{Derouillat2018}, by incorporating four key physics modules essential for accurately modeling quantum plasmas at the nanoscopic scale (\textcolor{blue}{Fig. \ref{fig:overview}B}).
(I) Initialisation of Fermi-Dirac distributed charge species – capturing the quantum statistics of electrons at low temperatures.
(II) Material boundary models – simulating the interaction of charge carriers with the material interfaces, crucial for condensed matter systems.
(III) Bound particle model for linear dispersion – enabling the simulation of material responses such as dielectric and plasmonic behavior in response to external fields.
(IV) Inclusion of massless Dirac fermions – to model materials like graphene, where electrons near the $K$-point exhibit relativistic behavior.
The modified versions of these codes with these modules are hosted publicly on Github~\cite{qepoch,qsmilei}. Each of these modules represents a significant advancement in adapting PIC to quantum plasma systems. After detailing these physics modules, we conclude with a discussion on the potential of combining them for simulating the dynamic, time-dependent physics of quantum plasmas in nanoscopic systems. This approach promises to enable more accurate and self-consistent simulations of condensed matter physics, with applications ranging from nanophotonics to structured quantum materials.

\section{Charge Carrier Momentum Distribution}
\begin{figure*}
    \centering
        \includegraphics[width=7.0in]{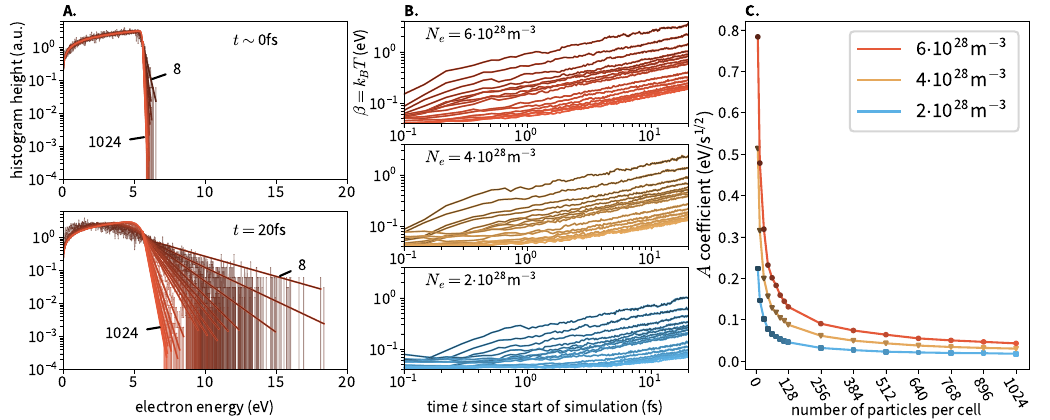}
        \label{fig:fddist}
    \caption{
    Relaxation of charged species initialized with a Fermi-Dirac (FD) distribution.
    (A) Evolution of electron macroparticles, initially initialized with FD distribution (top), relaxing toward a Maxwell-Boltzmann distribution after 20~fs. Each curve represents a different particle-per-cell weighting, with darker lines corresponding to lower and brighter lines to higher particle-per-cell values, with the minimum and maximum values labeled in each plot. 
    A Boltzmann factor $\sim\exp(-E/k_BT)$ is fitted to the tail of each histogram to quantify the deviation from the FD distribution. 
    (B) Evolution of the fitted temperature $k_BT$ for different simulation cases: \invmcube[2]{28} (blue), \invmcube[4]{28} (yellow), and \invmcube[6]{28} (red). 
    (C) Fitted coefficients to the square-root power law ($\beta = A\sqrt{t}$), showing the universal behavior observed in (B) as a function of the number of particles-per-cell. 
    }
    \label{fig:relax}
\end{figure*}

PIC codes typically model material with charged particle constituents following a Maxwell-Boltzmann (MB) distribution or, in some cases, a Maxwell-J{\"u}ttner distribution, especially when used for plasma simulations. The MB distribution is given by:
\begin{equation} 
    f_{\text{MB}}(\vec{v}) d^3v \propto \exp\left(-\frac{\frachalf m v^2}{k_B T}\right)d^3v, \label{eq:mbdist}
\end{equation}
where $m$ is the mass of the plasma species, 
$T$ is the temperature, and $k_B$ is the Boltzmann constant. This describes the velocity distribution of particles in a phase space volume $d^3v$ centered at velocity $\vec{v}$.

For modeling the electromagnetic response of condensed media in a PIC framework, it is more appropriate to initialize the momentum distribution of charged carriers using a Fermi-Dirac (FD) distribution, given the quantum nature of electrons in such systems \cite{kittelthermal}. The FD distribution is expressed as,
\begin{equation} 
    f_{\text{FD}}(\vec{v}) d^3v \propto \frac{1}{1+\exp\left(\frac{ \frachalf m v^2-\mu}{k_B T}\right)}d^3v, 
    \label{eq:fddist}
\end{equation}
where $\mu$ is the chemical potential, typically identified with the Fermi energy 
$E_F$ \cite{ashcroft_solid_1976}.

In practice, this can be implemented in PIC codes by using a custom loader to initialize macroparticles with velocities distributed according to the FD distribution in Eq. (\ref{eq:fddist}). 
Several open-source PIC codes, such as \textsc{epoch}, Smilei, and WarpX, include built-in support for such custom loaders, enabling the initialization of particles with non-Maxwellian distributions. In this case, the custom loader also handles the spatial distribution of particles to form the initial structure in the simulation.
More recently, \textsc{epoch} has introduced a direct method to specify the momentum distribution at the start of the simulation. The \verb|dist_fn| option allows users to set a custom momentum distribution for each species, alongside the specification of the momentum range for initialization. In our work, we choose a momentum range that ensures the initial electron velocities are initialized with the FD described in Eq.(\ref{eq:fddist}) with inputted Fermi energy and effective mass of electrons of the material under study.

Although macroparticles can be initialized with FD momentum distribution in PIC simulations, experience has shown that the momentum distribution of the charged species tends to relax to the MB distribution as the simulation progresses. This relaxation has been studied in other contexts, such as plasmas initialized in non-MB distributions~\cite{laietal2014}. 
This is not surprising, as particles in a PIC code are treated as free particles, and the MB distribution represents the equilibrium state of non-interacting free particles.
The key difference between the FD and MB distributions is that the FD distribution accounts for the Pauli exclusion principle, which prevents multiple particles from occupying the same quantum state. Enforcing the Pauli exclusion principle in a PIC code is challenging, as it involves discrete energy levels, while PIC codes model particles in the continuous space-time domain. In PIC simulations, energy is a derived quantity from particle momenta, and the occupation of discrete quantum states cannot be directly represented. Some attempts have been made to address this issue in PIC simulations, but a full computational solution to capture the Pauli exclusion principle remains elusive.
Therefore, for simulations of electromagnetic responses in plasmonic nanostructures, it is essential to run the simulation before the macroparticle species relax completely into an MB distribution, as this relaxation can affect the results of interest, particularly in the quantum regime.

Experience shows that while relaxation to the MB distribution is inevitable, initializing electron macroparticles with a higher particle-per-cell count helps maintain the FD distribution over longer simulation timescales, compared to simulations with lower particle-per-cell counts. To quantify the relaxation timescale and its dependence on particle resolution, we perform a study of bulk FD plasma across a range of electron densities $N_e$. Each simulation case involved a 2nm$\times$2nm$\times$2nm plasma cube (periodic boundary conditions were applied on all faces of the simulation box to model an infinite bulk material), initialized with an FD momentum distribution. The ions are fixed and the electrons are initialized randomly to specifically rule out other potential influences on the thermalization process (see \textcolor{blue}{SI-1} for more simulation details).  The temperature was set at 0.0375 eV, and the chemical potential was set to the Fermi energy 
$E_F = \frac{\hbar^2}{2 m} \left ( 3 \pi^2 N \right )^{3/2}$,
where $m$ is the mass of the charge carriers, 
$N$ is the carrier density, and 
$\hbar$ is the reduced Planck’s constant. 
For each electron density $N_e$, we varied the particle-per-cell count, with results shown in \textcolor{blue}{Fig.~\ref{fig:relax}}.

To quantify the relaxation dynamics, we analyze the high-energy tail of the electron macroparticle histogram at regular time intervals, fitting it to the functional form 
$\sim\exp(-E/\beta)$, where 
$E$ is the electron kinetic energy and 
$\beta$ is the temperature (related to the Boltzmann constant, $k_B T$). This allows us to track the evolution of the apparent temperature $\beta = k_B T$ over time.
In \textcolor{blue}{Fig. \ref{fig:relax}A}, we demonstrate the fitting procedure for a particular electron density $N_e=\invmcubemath[6]{28}$ at
$t\sim0$ and $t=20$ fs, in different simulations with varying numbers of particles per cell. 
We then plot the fitted $\beta$ for three different densities as a function of simulation time $t$ in \textcolor{blue}{Fig. \ref{fig:relax}B}, where we observe that the temperature follows an apparent square-root power law with respect to time: 
$\beta = A\sqrt{t}$.
The densities we selected are typical for plasmonic systems, assuming that each atom contributes one electron to the conduction band. For simple metals without impurities, these densities range from 
\invmcube[2]{28} to \invmcube[6]{28}. Across all simulation cases, the growth of $\beta$ was well-described by the power law 
$\beta = A\sqrt{t}$, with no significant differences in the numerical relaxation behavior regardless of the particle-per-cell count. 
However, the coefficient of the power law, 
$A$, varied between different cases, as shown in \textcolor{blue}{Fig.~\ref{fig:relax}C}.
Given that most metals have electron densities on the order of \invmcube{28} and similar Debye lengths, the numerical thermalization threshold was found to be roughly constant across all simulation cases. This explains why the coefficient 
$A$ converges around 128 particles per cell, even as the density varies. Consequently, this particle-per-cell threshold provides a practical lower limit for accurate PIC simulations of metals.

The study presented above focuses on the collisionless operation of our PIC codes, where relaxation is driven purely by statistical processes and occurs independently of any collisional model. In this context, relaxation to the MB distribution is expected as a result of the inherent statistical nature of the system. However, we observe that introducing collisions can accelerate the relaxation of a bulk plasma toward an MB distribution.
As a point of reference, the work by Wu et al. \cite{WuFDPIC2020} has implemented a method to maintain FD distribution over longer timescales when using a collisional module. They achieve this by histograms of macroparticles into energy bins, where Monte Carlo collisions are accepted only if the final energy bin is not occupied. A key insight of this approach is that the binning process is finite and only concerns energies below a cutoff above the Fermi energy 
$E_F$, as the FD distribution is expected to converge to an MB distribution well above 
$E_F$ in the classical limit.
However, this method is most effective when energy changes are relatively infrequent and occur in discrete events, making it more feasible to implement. It becomes more challenging to apply this approach at every time step when dealing with lower-energy macroparticles, where frequent updates to energy bins may be computationally demanding. While this technique could be useful for mitigating thermal relaxation in FD plasmas, it may require more sophisticated handling for efficient application over all simulation steps.
For the scope of our current work, where we are focused on the quantum plasma response of structures under electromagnetic radiation, the typical simulation sizes (on the order of microns) and short simulation durations (less than 100 fs) ensure that a sufficiently high particle-per-cell count is maintained. This particle resolution is adequate to preserve the FD distribution for the duration of the simulation, making the use of a collisional model unnecessary for our specific needs. Future work could explore implementing this procedure to improve the accuracy of FD plasma simulations over longer timescales or in more complex regimes. Ultimately, a complete PIC implementation to capture the Pauli exclusion principle is desirable.

\begin{figure*}[tbh]
    \centering
    \includegraphics[width=18.2cm]{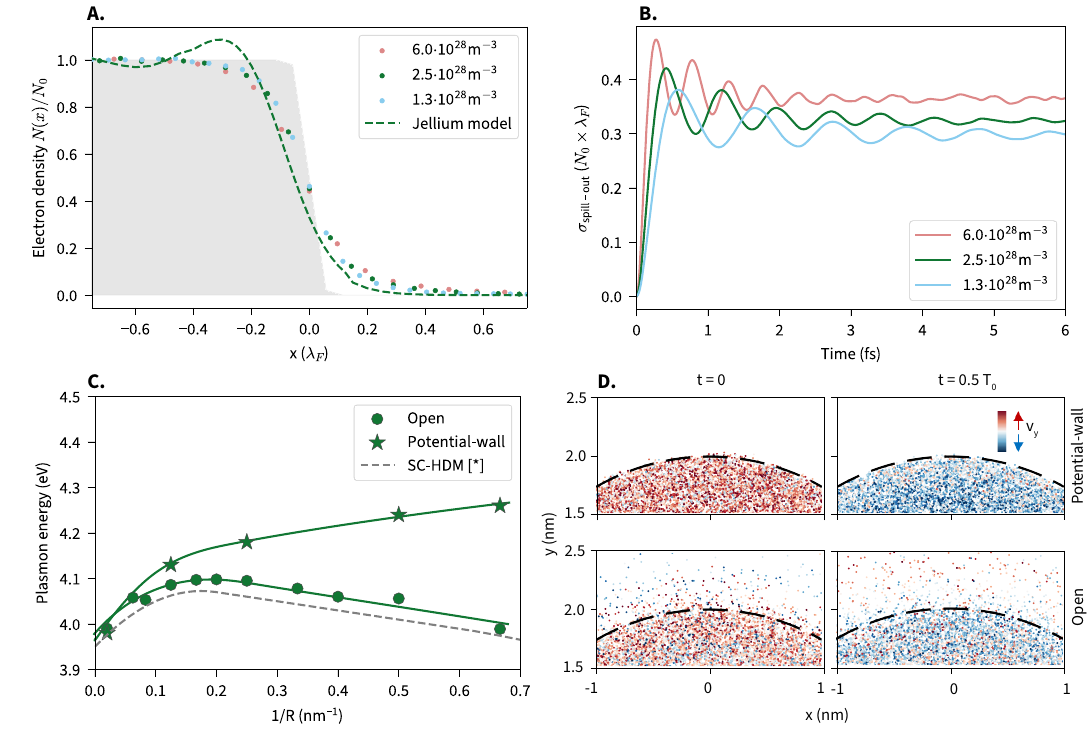}  
    \caption{Boundary conditions and charge spill-out.
    (A) Electron density distribution near the boundary of a semi-infinite material with varying bulk electron density $N_0$ and Wigner-Seitz radius ($r_s=3$ in red, $r_s=4$ in green, and $r_s=5$ in blue), compared with the Jellium model \cite{langkohn1970} at $r_s=4$ (Na). The solid grey region represents the ionic background.
    (B) Time evolution of the spill-out surface charge density, $\int_{0}^{\infty} N(x) \, dx$, for different $r_s$ values.
    (C-D) Tunable electrostatic boundary conditions in ultra-small Na nanocylinders, adapted from \cite{ding_particle_2020}. 
    (C) Plasmon energy versus cylinder radius $R$, comparing soft Open BC and hard potential-wall BC with the Self-consistent Hydrodynamic (SC-HDM) model \cite{toscano_resonance_2015}. 
    (D) Snapshots of the particle velocity distribution within a plasmon cycle for the potential-wall BC (top) and open BC (bottom). 
  }
    \label{fig:bcs}
\end{figure*}

\section{Material Boundary Conditions}\label{sec:bc}
In PIC simulations, material models are often used alongside particle modules to simulate plasmas interacting with devices or environments. These models typically define a spatially dependent dielectric constant or designate ``conductor cells'' to represent electrodes, allowing for complex voltage boundary conditions and current sources \cite{nieter2010, chacon2019}.
For nanophotonic quantum plasmas, the focus is on modeling the response of charge carriers ($e.g.$, electrons) using particle species. In addition to momentum space treatment, it is crucial to consider \emph{material boundary conditions}—the behavior of charged particles at the interface between the plasma and surrounding materials or free space. These conditions govern phenomena like reflection, transmission, and charge accumulation at material boundaries, which are key to accurately simulating electron emission and current flow in nanophotonic systems.

The simplest model for electron density at a metal surface is the Thomas-Fermi (TF) theory, foundational to modern density-functional theory. In TF theory, only Coulomb interactions and the electron gas's kinetic energy are considered. For a Fermi-Dirac electron gas, initialized with a step-like distribution at the metal surface, electrons with non-zero momentum spill out beyond the ionic boundary, forming an electric double layer. This causes the electron density to smoothly decrease from the metal into the vacuum over a length scale characterized by the Thomas-Fermi screening length. The resulting electrostatic barrier confines electrons within the material, while the kinetic energy of the gas creates a pressure gradient that balances the electrostatic potential. A similar model applies to degenerate plasmas in neutron stars and white dwarfs \cite{spruch_pedagogic_1991}.

In TF theory, the kinetic energy follows 
$T\propto N^{5/3}$, as dictated by the Pauli exclusion principle. The equilibrium electron density is reached by minimizing both the kinetic and electrostatic energies from the double layer. In PIC simulations, this energy minimization is also solved self-consistently to establish an equilibrium electron density at metal surfaces. In \textcolor{blue}{Figs.~\ref{fig:bcs}A, \ref{fig:bcs}B}, we show the equilibrium electron density distribution at metal surfaces $N(x)$ for varying bulk electron densities 
$N_0$, along with the transient electron spill-out. The spill-out electrons oscillate around the material boundary for several cycles, with the oscillation period determined by the surface plasmon energy, before stabilizing to the equilibrium distribution. The transient timescale is typically around 10 fs for normal metals. Despite not enforcing the Pauli exclusion principle, the relation 
$T\propto N^{1.56\pm0.1}$ still holds, close to TF theory's prediction (see \textcolor{blue}{SI-2} for reference).

However, TF theory and PIC models exclude exchange and correlation energies, leading to deviations in the electron density distribution compared to the jellium model, as shown in \textcolor{blue}{Fig.~\ref{fig:bcs}A}. The PIC model lacks spatial Friedel oscillations within the Fermi wavelength, which become more significant at low-density metals \cite{langkohn1970}. Nevertheless, both PIC and TF theory provide reasonable approximations of the electron density for metals with 
$r_s=2-6$, despite discrepancies due to the vanishing work function \cite{perdew_jellium_1988}.

To describe the work function in PIC, we use an external electrostatic field that does not interact with the Maxwell solver. This boundary condition models the effect of surface electron dynamics on optical nonlocal effects, as shown in \textcolor{blue}{Figs.~\ref{fig:bcs}C, \ref{fig:bcs}D}, adapted from our previous work \cite{ding_particle_2020}. 
The self-consistent electrostatic barrier confines electrons within the plasma, and the spatially varying electron density at the surface leads to a nonlocal optical response. Nonlocal effects arise from the smearing of induced charge (spill-out) within a finite length scale around the surface upon electromagnetic excitation. This causes a shift in the surface plasmon frequency when the nanostructure's characteristic length approaches the Thomas-Fermi screening length, around 0.5 nm for noble metals \cite{khurgin_ultimate_2015}. 
If the smear-out charge centroid is outside the ionic polarizable background (for simple metals), the surface plasmon is red-shifted; otherwise, it is blue-shifted (for noble metals). 
In the example shown, two boundary conditions—open and potential wall—are simulated in PIC, with predicted red-shifts and blue-shifts of the surface plasmon in ultra-small Na nanocylinders. These boundary conditions align with the self-consistent hydrodynamic model (SC-HDM) and hard-wall hydrodynamic model (HW-HDM) \cite{raza_unusual_2011,ciraci_hydrodynamic_2013,toscano_resonance_2015}. 
The red-shifted open boundary PIC case, including only electron kinetic and Coulomb interactions, matches well with the SC-HDM, which accounts for exchange and correlation effects.
The electrostatic boundary condition, therefore, can be further used to tailor the accurate work function on metal surfaces to include more quantum effects, such as tunneling and the generation of hot electrons \cite{wu_charge_2016}.

In PIC simulations, the precision of the external electrostatic field is constrained by the simulation cell size, as the field's extent depends on the interpolation order of the electric field. Small cell sizes, such as 0.05 nm in \textcolor{blue}{Figs.~\ref{fig:bcs}A, \ref{fig:bcs}B}, are necessary to accurately resolve the Thomas-Fermi screening length in normal metals, enabling precise external field resolution. However, for larger screening lengths, further reducing the cell size becomes both unnecessary and computationally impractical. To overcome this, conditional boundary conditions can be introduced to confine particles within the material boundary \cite{do2024}. These boundary conditions can incorporate scattering models, such as fully specular or diffusive scattering, to more accurately capture physics at the material boundary. We show that these scattering models significantly influence the optical nonlinearity of ballistic electrons \cite{do2024}. Future studies could extend this approach to model electron scattering along different crystallographic directions, for example, to capture enhanced plasmon damping at graphene's zigzag edges compared to its armchair edges \cite{thongrattanasiri_quantum_2012}.

\section{Modeling Dispersive Media Using Bound Species}
\begin{figure*}
    \centering
        \includegraphics[width=7.0in]{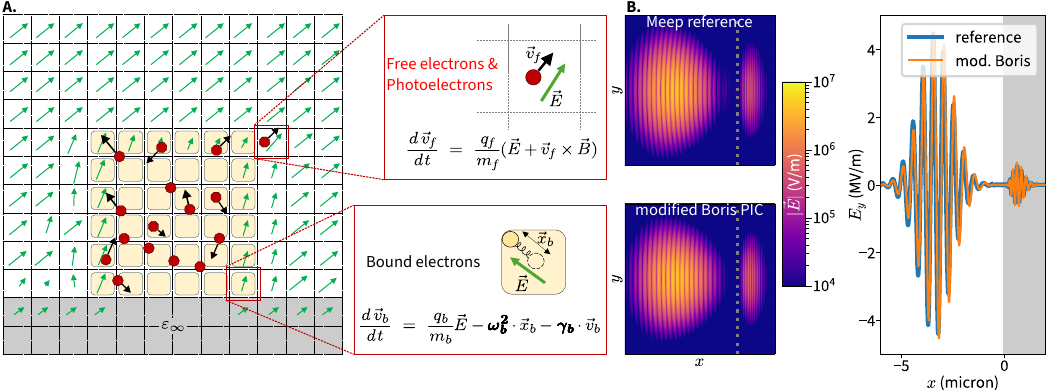}
    \caption{Bound species model for linear dispersive response. 
    (A) Schematic of the physics module combining different species in a single simulation: a constant dielectric function (gray cells), bound particles modeled by a damped harmonic oscillator (yellow cells), and free macroparticles (red). All species interact via electric and magnetic fields (green arrows).
    (B) Simulation example with a Lorentzian material: The left panels show a comparison between a \textsc{meep} simulation (top) and our code (bottom) for the reflection of a pulse incident on the material. The right panel shows a lineout along the laser axis, demonstrating excellent agreement over multiple orders of magnitude between the bound particle model and the FDTD reference case.
    }
    \label{fig:boundspecies}
\end{figure*}

PIC simulations are primarily designed to model plasma phenomena and thus typically exclude condensed matter effects where the optical response is dominated by interactions with bound electrons. However, some PIC codes do incorporate constant (but spatially varying) dielectric constants, often implemented as a cell quantity that modifies Maxwell's equations, particularly in the time derivative of 
$\vec{E}$ (curl of $\vec{B}$) during the field solver step. 
For explicit codes, where the 
$\vec{E}_{n+1}$ field is computed from the previous time step 
$\vec{E}_n$, we have that
\begin{equation}
    \vec{E}_{n+1}=\vec{E}_n+ c dt   \bm{\mathcal{E}}^{-1}(\vec{x})\cdot\left(\vec{\nabla}_s\times\vec{B}_{n+\frachalf}-\mu_0\vec{J}_{f,n}\right), \label{eq:consteps}
\end{equation}
where we omit spatial indexing for simplicity, as this depends on the specific interpolation scheme used. Here, $\vec{\nabla}_s\times$ represents the finite approximation of the curl in the chosen explicit scheme, $\bm{\mathcal{E}}^{-1}(\vec{x})$ is the inverse of a spatially varying dielectric tensor, and $\vec{J}_{f,n}$ is the current deposited by macroparticles in the given cell at the current time step. While incorporating a spatially varying, anisotropic dielectric tensor in PIC can introduce an anisotropic response, as implemented in both codes used in this work, it cannot capture the dispersive response typical of semi-metals and other condensed matter materials, which requires a frequency-dependent dielectric function.

To model a dispersive linear response in our PIC codes, we allow one or more specified charged particle species to be bound by a Lorentzian binding force that provides the desired electromagnetic response. This is achieved by adding an extra force term to the particle pusher. First, for the evolution of the velocity 
$\vec{v}_f$ for a free charged species 
$f$ with mass 
$m_f$ and charge 
$q_f$ is governed by the Lorentz force 
$\vec{F_f}$, given by:
\begin{equation} 
    \vec{F_f} (t) = \frac{d\vec{v}_f}{dt} = \qmr{f}  \left ( \vec{E} + \vec{v}_f \times \vec{B}  \right ) . \label{eq:tradlorentz}
\end{equation}
We now extend this equation to include a Hooke's law damped harmonic oscillator force for the dynamics of a bound charged species 
$b$, resulting in the following equation of motion:
\begin{equation} 
    \frac{d\vec{v}_b}{dt} = \frac{q_b}{m_b} \left ( \vec{E} + \vec{v}_b \times \vec{B}  \right )  - \bm{\omega_b^2} \cdot \vec{x_b} - \bm{\gamma_b} \cdot \vec{v_b} \label{eq:modlorentz},
\end{equation}
where $\bm{\omega_b^2}$ and $\bm{\gamma_b}$ are diagonal tensors that describe the binding force. For species $b$ with charge 
$q_b$ and mass $m_b$, irradiated by a monochromatic plane wave with angular frequency $\omega$, and assuming that the bound charges move much slower than the speed of light, the medium exhibits a Lorentzian-type resonance. This results in a linear dispersive response described by the relative dielectric tensor:
\begin{equation}
\varepsilon_b(\omega)/\varepsilon_0=I+\frac{q_b^2N_b}{m_b \varepsilon_0}
\frac{1}{\bm{\omega_b^2}-\omega^2-i\bm{\gamma_b}\omega}, \label{eq:lordielectric}
\end{equation}
where $I$ is the identity tensor and $\varepsilon_0$ is the permittivity of free space. 
Many materials that exhibit resonant dielectric functions can be approximated as a sum of Lorentzian species described by this relation, with $N_b$ representing the species' density and their relative contribution to the dielectric response. Moreover, for free species, the $\bm{\omega_b^2}$ and $\bm{\gamma_b}$ tensors can be set to zero to recover the original Lorentz force of Eq.~(\ref{eq:tradlorentz}).

To incorporate this model into the particle pusher of the PIC cycle, we modify the Boris pusher algorithm \cite{boris1970} to include the additional force terms. Specifically, for species 
$b$ with particle positions 
$\vec{x}_{b,n}$ defined on full time steps and velocities 
$\vec{v}_{b,n-\frachalf}$ on half steps, the position and velocity are advanced using the following scheme:
\begin{align} 
\label{eq:modborisscheme}
\vec{v}_{b,n-\frachalf}^- &= \vec{v}_{b,n-\frachalf} + \left ( \qmr{b} \vec{E}_n - \bm{\omega^2_b} \cdot \vec{x}_{b,n}  \right ) \nonumber \\
\vec{v}_{b,n+\frachalf}^+ &= \mathbf{R}(\bm{\gamma_b})\vec{v}_{b,n-\frachalf}^- \nonumber \\
\vec{v}_{b,n+\frachalf} &= \vec{v}_{b,n+\frachalf}^+ + \frac{dt}{2} \left ( \qmr{b} \vec{E}_n - \bm{\omega^2_b} \cdot \vec{x}_{b,n}  \right ) \nonumber \\
\vec{x}_{b,n+1} &= dt \vec{v}_{b,n+\frachalf},
\end{align}
where 
$\mathbf{R}(\bm{\gamma_b})$ is a rotation matrix that accounts for damping due to the velocity-dependent force $\bm{\gamma_b}\cdot \vec{v}_{b}$, and the other terms follow from the Lorentzian and electromagnetic forces. Further details on the terms are provided in the \textcolor{blue}{Appendix A.1, A.2}.

\textcolor{blue}{Fig.~\ref{fig:boundspecies}A} illustrates the inclusion of this capability as a particle species in the PIC framework. As shown schematically, different terms in the material response of a given medium, non-dispersive, constant dielectric response $\varepsilon_\infty$, dispersive response, and finally free charged particle current, are modeled self-consistently and simultaneously in our PIC codes. In this way, our PIC simulations can be applied well to systems of interest to nanophotonics and plasmonics. In particular, the electromagnetic response that can be attributed to charged particle motion over frequencies resolvable by the simulation time step is modeled by macroparticles native to PIC, albeit modified in the case of the bound particles. \textcolor{blue}{Fig.~\ref{fig:boundspecies}B} presents a basic benchmark, where a PIC simulation of a laser pulse interacting with a flat, bulk single Lorentzian material is compared to a similar FDTD \textsc{meep} simulation \cite{meepref} and finds almost exact agreement, validating the model for simulating dispersive material responses. These simulations are 2D3$v$, meaning they simulate two spatial dimensions and particles have all three dimensions of velocity for completeness of Maxwell's equations, but cannot move out of the simulation plane.

\subsection{Numerical Dispersion}
We now consider the stability of the modified Boris scheme incorporating the Lorentzian binding force. To analyze stability, we consider exponential solutions for the fields and currents of the form 
$\sim \exp( i \vec{k}\cdot \vec{x} - i \omega t)$ with wavevector 
$\vec{k}$ and frequency 
$\omega$. To properly account for interactions with the fields, we must solve Maxwell's curl equations, including the current density $\vec{J}$:
\begin{gather}
    \frac{1}{c} \partial_t \vec{E} =  c \vec{ \nabla} \times \vec{B} - \mu_0 \vec{J},\nonumber\\ \partial_t \vec{B} = - \vec{\nabla} \times \vec{E}.
\end{gather}
We use Yee's algorithm to discretize in space and time, where the electric field $\vec{E}_n$ is defined at full time steps, and the magnetic field 
$\vec{B}_n$ is defined at half time-steps. For simplicity, we neglect unbound charges, so Gauss's law is zero.

\begin{figure}
    \centering
    \includegraphics[width=3in]{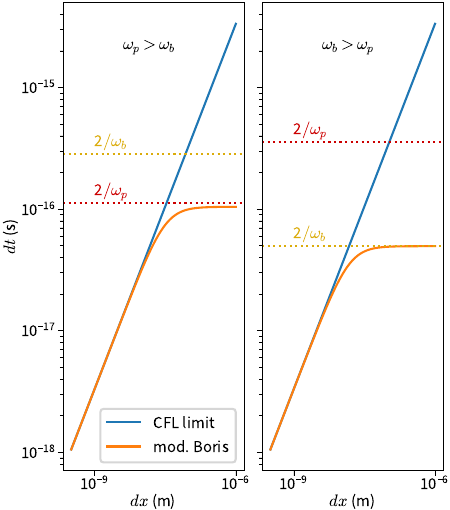}
    \caption{Time step limits for stability as a function of cell size $dx$.
    Left: Case where the plasma frequency $\omega_p$ is greater than the resonant frequency of the oscillator $\omega_b$.
    Right: Case where $\omega_b > \omega_p$. 
    The blue curve shows the well-known Courant-Friedrichs-Lewy (CFL) limit, while the orange curve represents the time-step limit for the modified Boris scheme, as given by Eq.~(\ref{eq:dtlimit}), in each case.
    }
    \label{fig:dtlimit}
\end{figure}

Focusing on a cell centered at the origin, the relation for the component $j=x,y,z$ of the electric field $E_n^j$ at time step $n$ becomes:
\begin{gather}
\left[  \sum_{l=x,y,z} \frac{\sin^2 ( k_l d x_l / 2)}{d x_l^2} - \frac{\sin^2(\omega d t /2)}{c^2 d t^2} \right] E^j_n \nonumber \\ =  \frac{i}{2 \varepsilon_0} J^j_n W(0,0,0) \frac{\sin(\omega d t/2)}{c d t}.
\end{gather}
Here, $W(0,0,0)$  is a weighting function depending on the chosen shape function for the simulation, 
and $J^j_n$ is current at time step $n$.
If $J^j_n e^{-i \omega d t / 2} = q_b N_b v^j_{n+\frachalf}$ for a density 
$N_b$ of resonators, we obtain a numerical dispersion relation for the system, after coupling with Eq.~(\ref{eq:modlorentz}):
\begin{gather}
    \left[  \sum_{l=x,y,z} \frac{\sin^2 ( k_l d x_l / 2)}{d x_l^2} - \frac{\sin^2(\omega d t /2)}{c^2 d t^2} \right] \bm{F}(\bm{\omega_b},\bm{\gamma_b})\nonumber \\ =  \frac{\omega_p^2}{c^2} \sin^2(\omega d t / 2) \bm{G}(k_j),\label{eq:numdispersion}
\end{gather}
where
\begin{equation}
    \bm{F}(\bm{\omega_b},\bm{\gamma_b}) = \bm{\omega_b^2} dt^2 - i dt \bm{\gamma_b} - 4 \sin^2(\omega d t/2) I.
    \label{eq:dispersion1}
\end{equation}
Here, $\omega_p^2=q_b^2 N_b / m_b \varepsilon_0 $ is the plasma frequency squared, and $\bm{G}$ is a matrix dependent on the shape function employed.  
For simplicity, we assume an isotropic resonance, so that 
$\bm{G}=I$, and one can show that the dispersion relation approaches the correct form in the limit 
$dt \rightarrow 0, dx \rightarrow 0$. Specifically, this results in a dielectric function of the form:
$\varepsilon(\omega)= 1 + \omega_p^2/(\omega_b^2-\omega^2 - i\gamma \omega)$,
which is equivalent to Eq.~(\ref{eq:lordielectric}), validating that the model behaves correctly for infinite resolution.
Considering the dispersion relation Eq.~(\ref{eq:dispersion1}) at the Nyquist frequency and wavenumbers where 
$\omega dt = k_l dx_l = \pi$, where aliasing is expected to be most prominent~\cite{Gordon2013} leads to the following constraint on the time step $dt$:
\begin{equation}
    d t < \frac{1}{\sqrt{\frac{1}{4}\omega_b^2 + \frac{1}{4}\omega_p^2 + c^2 \left ( \frac{1}{dx^2} + \frac{1}{dy^2} + \frac{1}{dz^2} \right )}},\label{eq:dtlimit}
\end{equation}
akin to the traditional Courant-Friedrichs-Levy (CFL) condition~\cite{CFL}. In \textcolor{blue}{Fig.~\ref{fig:dtlimit}}, we plot this limit in one spatial dimension for two cases: when 
$\omega_p>\omega_b$ and when 
$\omega_b>\omega_p$, as a function of the cell size $dx$. The time step limit saturates at twice the inverse of the highest frequency in the simulation, which is the key takeaway: to maintain stability, the time step must resolve the smallest time scale in the simulation. This result is simpler than the corresponding expression in~\cite{Gordon2013} and applies equally to other methods, such as Taflove's method~\cite{taflovehagness} and Varin's method~\cite{Varin2018}, which are arithmetically equivalent to this modified Boris scheme (see \textcolor{blue}{Appendix A.3}).

\begin{figure}
    \centering
    \includegraphics[width=3.0in]{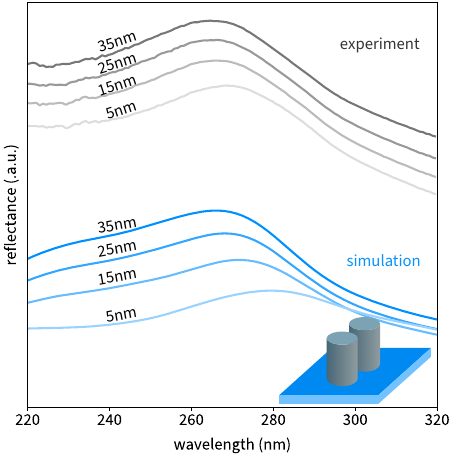}
    \caption{Test case of the dispersive bound particle model against experimental reflectivity of silicon nanostructures under ultraviolet irradiation. The experimental data from Dong et al. \cite{dong2019}, kindly provided by the authors, is compared with simulation results for 130-nm-tall, 70-nm-diameter silicon dimers with various gaps: 5, 15, 25, and 35 nm, irradiated by a Gaussian pulse with a broad bandwidth, ensuring nearly flat incident light across the reflectance range. The experimental and simulated plots are shifted vertically for clarity.
    }
    \label{fig:si4v}
\end{figure}

\subsection{Bound Species Test Cases}
To demonstrate the utility of this method, we present two simulation test cases. The first involves modeling silicon nanopillars under ultraviolet irradiation beyond its interband transitions~\cite{lautenschlager1987}. Experiments by Dong et al.~\cite{dong2019} have explored such structures in this wavelength range, where silicon exhibits a negative real dielectric function for wavelengths shorter than 300 nm, indicating potential plasmonic behavior. Dong et al. observed mode hybridization in dimer nanostructures by examining their spectral reflectivity as a function of the separation gap.

Using Eq.~(\ref{eq:lordielectric}), we fit four Lorentzian resonances to the measured dielectric function from that paper (see \textcolor{blue}{SI-3} for details) and computed the spectral reflectivity of dimer structures matching those in the experimental setup. The results, shown in \textcolor{blue}{Fig.~\ref{fig:si4v}}, exhibit excellent agreement with the experimental shift in the reflectance peak as a function of gap size. However, our simulations predict a more pronounced shift at the smallest gap (5 nm), while the experimental data displays a more subtle shift across the different gap values tested.

As a further demonstration of the bound particle model, we consider radiation emission due to electron transitions through a material. In this scenario, Cherenkov radiation (CR) or Transition Radiation (TR)~\cite{Ginzburg1996} is typically expected. PIC simulations are particularly suited for modeling TR, as the process is inherently time-dependent, and PIC provides self-consistent macroparticles to track the dynamics of accelerated charge bunches that generate radiation.

\begin{figure}
    \centering
    \includegraphics[width=3in]{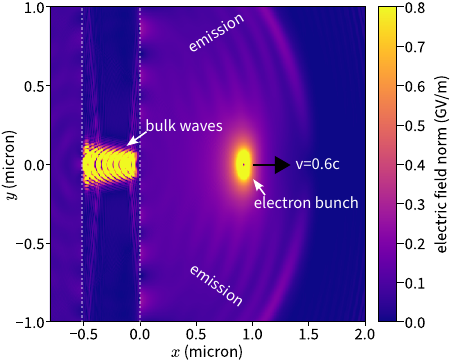}
    \caption{Test case for the transition of an electron bunch moving at $0.6c$ through a dispersive dielectric barrier with radiation emission. The bulk bound waves are shown, continuing to oscillate after the electron bunch has exited the material, indicating the persistent emission of radiation.}
    \label{fig:tr}
\end{figure}

Conventional PIC simulations often model the medium through which the electron moves with a constant dielectric, as described by Eq.~(\ref{eq:consteps}). However, as shown in the spatial field distribution in \textcolor{blue}{Fig.~\ref{fig:tr}}, using bound species to represent the medium not only captures its dispersive electromagnetic response but also allows investigation of the medium's dynamics. The simulations in \textcolor{blue}{Fig.~\ref{fig:tr}} feature an electron bunch traveling at $0.6c$ ($c$ represents the speed of light in vacuum) through a medium characterized by a single Lorentzian resonance. At the time step shown, the electron bunch has already passed through the medium, but the medium continues to oscillate and emit radiation due to the kinetic energy imparted by the bunch.

Such dynamics cannot be observed in simulations where the medium is modeled with a simple dielectric constant (see \textcolor{blue}{SI-4}). However, by modeling the medium’s electrons as macroparticles, as in the presented method, these effects are naturally captured, enabling a deeper understanding of the medium's response.

\begin{figure*}
    \centering
    \includegraphics[width=7.0in]{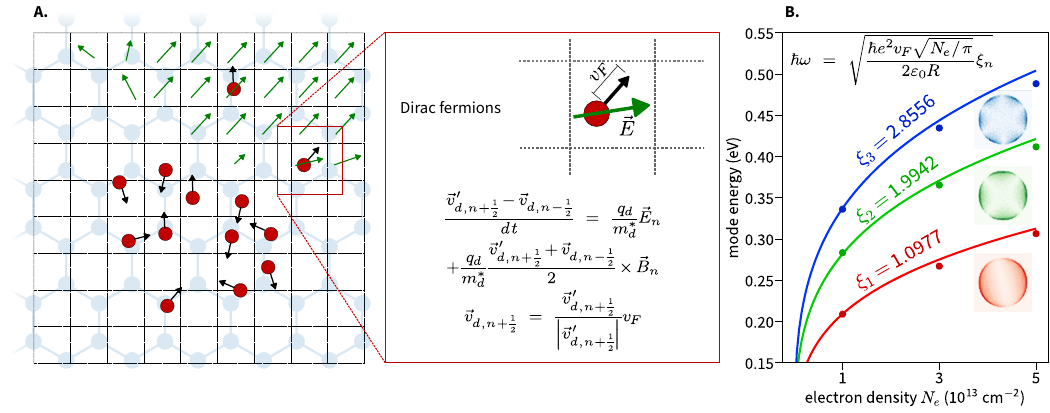}
    \caption{Inclusion of massless Dirac fermions to model graphene-like materials. 
    (A) Schematic of the particle motion near the $K$-point in graphene, where electrons are pushed by the Lorentz force (right inset), and their velocities are renormalized to the Fermi velocity $v_F$.
    (B) Test case of whispering-gallery modes in a graphene disc, with the observed frequencies matching analytic theoretical calculations as a function of carrier density, as shown in the inset equation.} 
    \label{fig:grapheneover}
\end{figure*}

\section{Dirac Electrons in 2D Materials}
The use of free electron macroparticles in PIC simulations is well-suited for modeling the dynamics and response of conduction band electrons in quantum plasmas, particularly when these electrons exhibit a parabolic dispersion relation for their energy $E$:
\begin{equation}
E = E_0+ \frac{\hbar k^2}{2 m^*},
\end{equation}
where $E_0$ is the zero-point energy, $m^*$ is the effective electron mass, and $\vec{p}=\hbar \vec{k}$ is the quantized electron momentum. 
This model is adequate for many metallic materials, as demonstrated in our previous works~\cite{ding_particle_2020, do2021}.
However, for materials with linear dispersion near a Dirac point in their band structure, such as graphene, a more nuanced treatment is required. Near the Dirac point, electrons follow the following linear dispersion relation:
\begin{equation}
E={\hbar}{v_{F}}k,
\end{equation}
which results in a constant group velocity of the electron wave packet at Fermi velocity $\frac{1}{\hbar}\partial_{k}E=v_F$ and a zero effective mass of electrons $m^{*}= \frac{1}{\hbar^2}\partial^{2}_{k}E(k)=0$ \cite{castro_neto_electronic_2009}. To describe the massless feature of Dirac electrons in PIC, we can employ a non-isotropic inverse mass tensor:
\begin{equation}
M^{*-1}_{ij}=\frac{1}{\hbar^2}\partial_{k^i}\partial_{k^j}E(k).
\end{equation}
In this section, we adopt relativistic covariant notation, where Latin indices represent spatial coordinates and Greek indices denote space-time coordinates. Repeated indices imply summation over spatial or space-time dimensions, and we use the standard Einstein summation convention. In the case of 2D electron gas in graphene, the inverse mass tensor has two eigenvalues$ \{0,v_{F}/p\}$  corresponding to the momentum eigenvector parallel and perpendicular to the direction of the momentum vector $\vec{p}$. This means that effective mass is infinite in the longitudinal direction with $\vec{p}$ and finite at $m_{d}=p/v_{F}$ in the transverse direction.

In this context, a natural description for such species is to treat them as following the Dirac equation for massless fermions, with the speed of light 
$c$ replaced by the Fermi velocity 
$v_F$:
\begin{equation}
i \hbar \gamma^\mu \partial_\mu' \psi(x,t)  =  0 \label{eq:diracelectrons}
\end{equation}
where $\gamma^\mu$ are the gamma matrices, defined in terms of the Pauli matrices $\sigma^i$,
such that  
$\gamma^i \equiv \begin{pmatrix}
    0 & \sigma^i \\ \sigma^i &0\
\end{pmatrix}$ and $\gamma^0 = \begin{pmatrix}
    I &  0 \\
    0 & -I
\end{pmatrix}.$ 
In this formulation, the covariant derivative 
$\partial'^\mu=(\partial'^0,\vec{\nabla})$ is redefined to account for the Fermi velocity $v_F$ instead of the speed of light in the time derivative:
\begin{equation}
    \partial'^0 \equiv \frac{1}{v_F} \frac{\partial}{\partial t}
\end{equation}
This redefinition ensures the correct treatment of the relativistic effects at the Dirac point, where the electron dynamics are governed by the Fermi velocity rather than the speed of light.

The fact that Dirac electrons behave as massless fermions moving at a constant speed, specifically the Fermi velocity $v_F$, motivates modeling these carriers with macroparticles whose speeds are capped at $v_F$, while still allowing them to be influenced by electromagnetic fields. To achieve this, we have modified the traditional Boris pusher in PIC simulations to accommodate the unique dynamics of Dirac fermions. This approach is schematically illustrated in \textcolor{blue}{Fig.~\ref{fig:grapheneover}A}.

Starting from the conventional Boris scheme in Eq.~(\ref{eq:borisscheme}), the velocity update procedure is modified as follows: after each iteration of the Lorentz force push, the particle's velocity is renormalized to the Fermi velocity. The inputted particle mass is the transverse mass $m_{d}=p/v_{F}$, while the velocity renormalization ensures infinite longitudinal effective mass.
Explicitly, the velocity 
$\vec{v}_{d,n-\frachalf}$ at the previous half-step is updated by the Lorentz force, yielding an intermediate velocity $\vec{v}_{d,n+\frachalf}'$. This intermediate velocity is then normalized to the Fermi velocity, producing the final velocity for the next half-step, 
$\vec{v}_{d,n+\frachalf}$, before advancing the position:
\begin{align}
\vec{v}_{d,n-\frachalf}^- &= \vec{v}_{d,n-\frachalf} + \frac{dt}{2} \qmr{d} \vec{E}_n \nonumber\\
\vec{v}_{d,n+\frachalf}^+ &= \mathbf{R}\vec{v}_{d,n-\frachalf}^- \nonumber\\
\vec{v}_{d,n+\frachalf}' &= \vec{v}_{d,n+\frachalf}^+ + \frac{dt}{2} \qmr{d}
\vec{E}_n \nonumber\\
\vec{v}_{d,n+\frachalf} &= v_F \frac{\vec{v}_{d,n+\frachalf}'} {\left | \vec{v}_{d,n+\frachalf}'\right| } \nonumber\\
\vec{x}_{d,n+1} &= \vec{x}_{d,n} + dt \vec{v}_{d,n+\frachalf}
\end{align}

To validate this method for modeling Dirac fermions in graphene, we performed benchmark simulations of graphene nano-disks with a 50 nm diameter, examining their oscillatory modes. These simulations considered three different Dirac electron densities: 
$\topowmath[1]{13}$~cm$^{-2}$, $\topowmath[3]{13}$~cm$^{-2}$, and $\topowmath[5]{13}$~cm$^{-2}$. The graphene nanodisks are excited with a fast-moving $1000e$ at $0.5c$ along the $z$-direction perpendicular to the nanodisks. The Dirac 2D electron gas is formed by charged particles with zero momentum in the $z$-direction. By applying Fourier analysis, we extracted the frequencies of the first three resonance modes of the graphene disks, as shown in 
\textcolor{blue}{Fig. \ref{fig:grapheneover}B}. The results closely match the analytical predictions found in the literature \cite{christensen_classical_2017}, demonstrating the reliability of the model.

In this example, we use an averaged transverse effective mass $m_d=p_{F}/2v_{F}$ for all electrons, with $p_{F}$ being the momentum of the electrons at the Fermi level. This yields the correct plasmon mass per electron $m_{\text{plasmon}}=p_{F}/v_{F}$ \cite{yoon_measurement_2014}, explaining the good match in plasmon resonance frequencies. The currently implemented model can describe the collective effect of the Dirac electron gas well. However, as the electrons have fixed Fermi velocity and effective mass, they effectively do not gain energy throughout the simulation and the temperature of the electron gas is maintained at $0$ K. A non-zero electron temperature can be implemented in future work. 
This new model of Dirac electrons has been successfully applied to a study investigating resonances in shaped graphene bow-tie structures, revealing a previously unobserved nonlinearity arising from surface scattering \cite{do2024}.

\section{Coupling of Modules and Prospects}

In the previous sections, we have detailed a number of key physics modules incorporated into modified versions of two popular, open-source PIC codes, \textsc{epoch} and Smilei. These modules are essential for modeling quantum plasmas in the low-temperature, condensed matter regime, with a particular emphasis on nanoscopic electromagnetic responses. We now highlight that one of the most promising aspects of this methodology for modeling condensed matter in the nanoscopic, cold regime is its ability to couple existing modules in PIC codes to the physics discussed in this article.
For example, Ding et al. (2020) \cite{ding_particle_2020} showed that the open boundary condition discussed in \textcolor{blue}{Section \ref{sec:bc}} yields the best match to theory for the spill-out phenomenon, as shown in \textcolor{blue}{Fig. \ref{fig:bcs}}. This represents one of the simplest yet most important examples of coupling the natural diffusive particle dynamics modeled in PIC simulations with material boundaries, thus providing insights into quantum plasma physics. Similarly, Do et al. (2021) \cite{do2021} demonstrated how the inclusion of the Perez collision model \cite{perez2012} in PIC codes, along with particles initialized in a Fermi-Dirac distribution, provided insights into the damping of modes in gold nanorod structures under irradiation. In both cases, the integration of the physical models discussed in this article with the existing capabilities of PIC codes has enhanced our understanding of nanoscopic material dynamics—dynamics that are typically not accounted for in traditional FDTD or FEM codes without significant modifications.

While we are optimistic about the potential of this method for modeling quantum plasmas, it is important to note one of its key limitations: it remains a semi-classical approach. Notably, it lacks a comprehensive treatment of energy levels and the emission of incoherent photons, particularly those with wavelengths smaller than the grid resolution ($i.e.$, smaller than the cell size or those that cannot be resolved by the Yee grid). While \textsc{epoch} does include a photon model, it is focused on high-energy photons resulting from non-linear quantum electrodynamic interactions, not the lower-frequency, radiative photons typically associated with energy losses in condensed matter. Future work will need to address these shortcomings by developing methods to simulate these effects more effectively in the context of PIC codes, thus enhancing their utility for modeling condensed matter media.

One of the most exciting prospects for the approach outlined in this article is the ability to model inherently time-dependent, high-energy-density physics (HEDP). 
As demonstrated in \textcolor{blue}{Fig.~\ref{fig:tr}} and other studies \cite{Gordon2013}, PIC is a powerful tool for simulating free-electron radiation generation schemes. It is already widely used to model high-intensity laser-plasma interactions~\cite{strehlow2024, Willim2024, Knight2024, ngirmang2020} and it is well-suited to the study of transition radiation, Cherenkov radiation, Smith-Purcell radiation, and electron loss spectroscopy, among other electron-matter interactions that involve light emission. In the HEDP domain, where PIC is already extensively used to model relativistic laser-plasma interactions, the incorporation of condensed matter techniques enables simulations of intense, short-pulse laser interactions in cases where the material remains less ionized than in typical high-intensity scenarios. This makes it ideal for studying transient, temporally dependent material responses, such as those observed in laser damage to dielectric materials or nano-machining using high-intensity lasers~\cite{Shcherbakov2023}.

By combining the strengths of quantum plasma modeling with traditional PIC methods, we can better explore the ``particle'' side of quantum plasmas—particularly in regimes that lie farther from the quantum plasma threshold. This opens up novel possibilities for investigating the physics of plasmas in a regime that has remained relatively under explored in computational plasma physics. Thus, as the field progresses, the integration of quantum plasma physics into PIC codes holds great promise for advancing our understanding of both the fundamental and applied aspects of condensed matter interactions.

\section{Conclusion}
In conclusion, we have integrated four key physics modules into the \textsc{epoch} and Smilei open-source PIC codes: 
(I) Fermi-Dirac distributed charged species, (II) boundary conditions for macroparticles at material edges, (III) a bound-particle model for linear dispersive responses, and (IV) a framework for modeling massless Dirac fermions, ideal for simulating graphene-like materials. These modules provide valuable insights into the time-domain simulation of electromagnetic interactions with quantum plasmas, enabling more dynamic and accurate modeling of light-matter interactions at the nanoscale.

This work bridges the gap between traditional electromagnetic simulations and the complex dynamics of quantum systems by integrating physics models into a single computational framework. It significantly advances our understanding of nanophotonics, particularly in plasmonics and photon confinement, and offers broad potential for fields involving light and quantum plasma interactions. This includes applications in advanced quantum materials, high-energy-density plasmas, optoelectronics, polariton physics, petahertz electronics, and symmetry-fusion topological photonics, opening new avenues for both fundamental and applied research.

\section{Acknowledgements}
This work was supported by the
Ministry of Education Singapore (Grant MOE-T2EP50223-0001), the National Research Foundation Singapore (Grants NRF2021-QEP2-02-P03, NRF2021-QEP2-03-P09, NRF-CRP26-2021-0004), and the Singapore University of Technology and Design (Kickstarter Initiatives SKI 2021-02-14, SKI 2021-04-12).
G. Liu acknowledges support from the Ph.D. RSS. H. T.B. Do and M. Bosman acknowledge support from the Singapore Ministry of Education (Grant MOE-T2EP50122-0016).

\section{CRediT Statement}
This section summarizes the contributions of each author according to the CRediT scheme~\cite{Credit}. Gregory K. Ngirmang and Hue T. B. Do are co-first authors, contributing equally to: Conceptualization, Data curation, Formal analysis, Investigation, Methodology, Resources, Software, Validation, Visualization, and Writing (Original Draft and Review \& Editing).
Guangxin Liu contributed to: Conceptualization, Investigation, and Writing (Original Draft).
Michel Bosman contributed to: Funding acquisition, Project administration, Resources, Supervision, and Writing (Review \& Editing).
Lin Wu contributed to: Conceptualization, Funding acquisition, Investigation, Methodology, Project administration, Resources, Supervision, Writing (Original Draft and Review \& Editing).

\section{Declaration of Competing Interests}
The authors declare that they have no competing financial, professional, or personal interests that could influence or bias the work presented in this article.

\newpage
\appendix

\section{Derivation of Modified Boris Scheme}

To derive the modified Boris scheme in Eq.~(\ref{eq:modborisscheme}), we begin by revisiting the traditional Boris scheme, which is commonly employed in PIC codes for the explicit evolution of macroparticles under the Lorentz force, as described by Eq. ~(\ref{eq:tradlorentz}). The standard Boris pusher is a well-established method for updating the velocity and position of particles in electromagnetic fields in a manner that preserves the accuracy and stability of the integration process. By modifying this standard approach to incorporate additional forces, such as those arising from binding interactions or material-specific responses, we can extend its applicability to more complex scenarios, such as bound-electron species or materials exhibiting dispersive behavior. This process involves introducing the appropriate force terms and updating the velocity and position steps accordingly while maintaining the particle's dynamics in a self-consistent and computationally efficient manner.

\subsection{Traditional Boris Scheme}
We consider the traditional Lorentz force 
$\vec{F_f}(t)$ acting on a charged species $f$ with charge $q_f$ and mass $m_f$ as given by Eq.~(\ref{eq:tradlorentz}). The widely used Boris scheme \cite{boris1970} solves the Lorentz force for macroparticles, generating discrete time series of macroparticle positions 
$\vec{x}_{f,n}$ and velocities 
$\vec{v}_{f,n+\frachalf}$, with a uniform time step $dt$. To derive the scheme, we discretize Eq.~(\ref{eq:tradlorentz}) as a second-order finite-difference equation:
\begin{gather}
    \frac{\vec{v}_{f,n+\frachalf} - \vec{v}_{f,n-\frachalf}}{d t} = \nonumber \\
    \frac{q_f}{m_f} \left ( \vec{E}_n + \frac{\vec{v}_{f,n+\frachalf} + \vec{v}_{f,n-\frachalf}}{2} \times \vec{B}_n \right ), \label{eq:discrlorentz}
\end{gather}
\begin{equation}
    \vec{x}_{f,n+1} = \vec{x}_{f,n} + d t \vec{v}_{f,n + \frachalf}. \label{eq:discrvel}
\end{equation}
Here, $\vec{E}_n$ and $\vec{B}_n$ are the fields interpolated onto the particle position at the whole time step $n$.
The Boris scheme steps this equation by introducing an intermediate velocity step as:
\begin{equation}
    \vec{v}_{f,n \pm \frachalf}^{\pm} = \vec{v}_{f,n\pm \frachalf} \mp \frachalf \frac{q_f}{m_f} \vec{E}_n dt, \label{eq:borisvpm}
\end{equation} 
transforming Eq.~(\ref{eq:discrlorentz}) into:
\begin{equation}
    \frac{\vec{v}_{f,n+\frachalf}^{+}-\vec{v}_{f,n-\frachalf}^{-}}{d t} = \qmr{f} \left ( \vec{v}_{f,n+\frachalf}^{+} + \vec{v}_{f,n-\frachalf}^{-} \right ) \times \vec{B}_n, \label{eq:borisrot}
\end{equation}
which can be solved to find $\vec{v}_{f,n + \frachalf}^{+}$.  A second electric field push is then applied to yield the updated velocity $\vec{v}_{a,n + \frachalf}$. 

The solution for $\vec{v}_{a,n+\frachalf}^+=(v_x^+,v_y^+,v_z^+)^T$ in terms of $\vec{v}_{a,n-\frachalf}^-=(v_x^-,v_y^-,v_z^-)^T$ is given by:
\begin{align}
    \vec{v}_{f,n+\frachalf}^+ &= \mathbf{R}  \vec{v}_{f,n-\frachalf}^-, \nonumber \\
    \mathbf{R} &= \frac{I (1-|b_{f,n}|^2) + 2 \left (\vec{b}_{f,n} \otimes \vec{b}_{f,n} \right ) + 2 \bm{\varepsilon}(\vec{b}_{f,n})}{1 + |b_{f,n}|^2}, \label{eq:borisrotsol}
\end{align} 
where $\mathbf{R}$ is a matrix representing the rotation of the velocities due to the magnetic field $\vec{B_n}$, $I$ is the identity matrix, $\otimes$ is the tensor product, $\vec{b}_{f,n}= dt \vec{B}_n q_f / (2 m_f)$, and $\bm{\varepsilon}(\vec{v})$ being the antisymmetric matrix representing the cross product with the vector $\vec{v}$:
\begin{equation*}
    \bm{\varepsilon}(\vec{v}) = \begin{pmatrix}
     0    & v_z  & -v_y \\
     -v_z &  0   &  v_x \\
     v_y  & -v_x &    0 \\
    \end{pmatrix}\text{, for } \vec{v} = \begin{pmatrix}
    v_x \\ v_y \\ v_z \end{pmatrix}.
\end{equation*} 
The position is then updated using  Eq.~(\ref{eq:discrvel}). 

Thus, the Boris scheme computes the next time step position $\vec{x}_{f,n+1}$ and velocity $\vec{v}_{f,n+\frachalf}$ from the current position 
$\vec{x}_{f,n}$ and velocity $\vec{v}_{f,n-\frachalf}$ as:
\begin{align}
\label{eq:borisscheme}
\vec{v}_{f,n-\frachalf}^- &= \vec{v}_{f,n-\frachalf} + \frac{dt}{2} \qmr{f} \vec{E}_n \nonumber\\
\vec{v}_{f,n+\frachalf}^+ &= \mathbf{R}\vec{v}_{f,n-\frachalf}^- \nonumber \\
\vec{v}_{f,n+\frachalf} &= \vec{v}_{f,n+\frachalf}^+ + \frac{dt}{2} \qmr{f} \vec{E}_n \nonumber\\ 
\vec{x}_{f,n+1} &= dt \vec{v}_{f,n+\frachalf}.
\end{align}
While the Boris scheme is not symplectic, Qin et al.~\cite{Qin2013} demonstrated that it preserves phase space, which accounts for its stability and accuracy in practice.

\subsection{Modified Boris Scheme}
We consider the discretization and solution of the Lorentz force with an additional damped harmonic oscillator force, as described in Eq.~(\ref{eq:modlorentz}). A second-order accurate discretization of this equation is:
\begin{gather}
\frac{\vec{v}_{b,n+\frachalf} - \vec{v}_{b,n-\frachalf}}{d t} = \qmr{b} \vec{E}_n - \bm{\omega_b^2} \cdot \vec{x}_{b,n} \nonumber \\- \frachalf \bm{\gamma_b} \cdot \left ( \vec{v}_{a,n+\frachalf} + \vec{v}_{a,n-\frachalf} \right ),\label{eq:discrmodlorentz}
\end{gather}
We first neglect the magnetic field and apply the same half-push treatment to the Hooke's law term as in Eq.~(\ref{eq:borisvpm}), re-defining the intermediate velocities similar to the Boris push:
\begin{equation}
    \vec{v}_{b,n \pm \frachalf}^{\pm} = \vec{v}_{b,n\pm \frachalf} \mp \frac{dt}{2} \left ( \frac{q_b}{m_b} \vec{E}_n - \bm{\omega^2_b} \cdot \vec{x}_{b,n} \right ).
    \label{eq:modborisvpm}
\end{equation} 
This substitution transforms Eq.~(\ref{eq:discrmodlorentz}) into a rotation-like equation similar to Eq.~(\ref{eq:borisrot}):
\begin{equation}
       \frac{\vec{v}_{b,n+\frachalf}^{+}-\vec{v}_{b,n-\frachalf}^{-}}{d t} = \qmr {b} \bm{\gamma_b} \cdot \left ( \vec{v}_{b,n+\frachalf}^{+} + \vec{v}_{b,n-\frachalf}^{-} \right ), \label{eq:modborisrot}
\end{equation}
which has a solution:
\begin{align}
\vec{v}_{b,n+\frachalf}^+ &= \mathbf{R}(\bm{\gamma_b})  \vec{v}_{b,n-\frachalf}^-, \nonumber \\
    \mathbf{R}(\bm{\gamma_b}) &= \begin{pmatrix}
        \frac{1-\frachalf \gamma_{b,x} dt }{1+\frachalf \gamma_{b,x} dt} & 0 & 0 \\
        0 & \frac{1-\frachalf \gamma_{b,y} dt}{1+\frachalf \gamma_{b,y} dt} & 0 \\
        0 & 0 & \frac{1-\frachalf \gamma_{b,z} dt}{1+\frachalf \gamma_{b,z} dt} \\ 
    \end{pmatrix},
\label{eq:modborisrotsol}
\end{align}
for the diagonal dampening matrix $\bm{\gamma_b}=\text{diag}(\gamma_x,\gamma_y,\gamma_z)$. 
Thus, the modified Boris scheme for bound species is:
\begin{align} 
\label{eq:modborisscheme2}
\vec{v}_{b,n-\frachalf}^- &= \vec{v}_{b,n-\frachalf} + \left ( \qmr{b} \vec{E}_n - \bm{\omega^2_b} \cdot \vec{x}_{b,n}  \right ) \nonumber \\
\vec{v}_{b,n+\frachalf}^+ &= \mathbf{R}(\bm{\gamma_b})\vec{v}_{b,n-\frachalf}^- \nonumber \\
\vec{v}_{b,n+\frachalf} &= \vec{v}_{b,n+\frachalf}^+ + \frac{dt}{2} \left ( \qmr{b} \vec{E}_n - \bm{\omega^2_b} \cdot \vec{x}_{b,n}  \right ) \nonumber \\
\vec{x}_{b,n+1} &= dt \vec{v}_{b,n+\frachalf}.
\end{align}

For non-bound particles, setting 
$\bm{\gamma_b}=0$ and $\omega_b=0$ restores the original Boris scheme, while the magnetic field force must be reintroduced, requiring a weighing factor to be implemented in the solver to enable or disable the magnetic field term in the pusher.

In the case where the magnetic field is non-zero, we utilize the same intermediate velocities in Eq.~(\ref{eq:modborisvpm}) and obtain the rotation equation:
\begin{equation}
\frac{\vec{v}_{b,n+\frachalf}^+-\vec{v}_{b,n-\frachalf}^-}{dt} =\qmr{b}(\varepsilon(\vec{B}_n)-\bm{\gamma_b})\cdot \frac{\vec{v}_{b,n+\frachalf}^++\vec{v}_{b,n-\frachalf}^-}{2}.
\label{eq:modborisBrot}
\end{equation}
Solving this equation requires the following, somewhat more complex rotation matrix:
\begin{gather}
\mathbf{R}=\left( \frac{I-\halfdt\bm{\gamma_b} }{I+\halfdt\bm{\gamma_b}}\text{Det}(I+\halfdt\bm{\gamma_b})+\bm{\varepsilon}((I+\halfdt\bm{\gamma_b})\cdot\vec{b}_{b,n}) \nonumber \right.\\\left.+\vec{b}_{b,n}\otimes\vec{b}_{b,n} \right)/\text{Det}(L_-) \\
\text{Det}(L_-)=\text{Det}(I+\frac{dt}{2}\bm{\gamma_b})+\vec{b}_{b,n}\cdot((I+\halfdt\bm{\gamma_b})\cdot\vec{b}_{b,n}),\\
\vec{b}_{b,n} = \halfdt\qmr{b}\vec{B}_n.
\end{gather}
This matrix can replace the rotation matrix in Eq.~(\ref{eq:modborisscheme2}) to model gyroscopic media. Thus, this modified Boris scheme can be used to model both bound and non-bound species, including magnetic effects, in PIC simulations of gyroscopic media.

\subsection{Comparison to Other Works}
We note that the derived equations are functionally equivalent to those presented by Varin et al.~\cite{Varin2018}, specifically the stepped form in Eq.~8(a-b) of their work. The primary distinction in our approach lies in the use of a particle species to model the dipole response, rather than a cell quantity for the dipole polarization, as employed in their method and other similar works by these authors.
Furthermore, by substituting Eq.~(\ref{eq:discrvel}) directly into Eq.~(\ref{eq:discrmodlorentz}), we obtain the following expression:
\begin{gather}
\frac{\vec{x}_{b,n+1} - 2 \vec{x}_{b,n} + \vec{x}_{b,n-1}}{d t} \nonumber \\ = \frac{q_a}{m_a} \vec{E}_n - \bm{\omega_b^2} \cdot \vec{x}_{b,n} - \bm{\gamma_b}\cdot \frac{\vec{x}_{a,n+1} - \vec{x}_{a,n-1}}{2 d t},\label{eq:taflove}
\end{gather}
which mirrors the formulation introduced by Taflove~\cite{taflovehagness} and is commonly found in FDTD codes. This similarity accounts for the close agreement observed in \textcolor{blue}{Fig.~\ref{fig:boundspecies}}. Thus, the modified Boris scheme is algebraically equivalent to Taflove's solution, with the key difference being the way the velocity (or, in the case of Varin et al., the current) is implemented as a macroparticle species instead of being a cell quantity. Moreover, while Taflove's method requires storing the resonance displacement from the previous time step (\emph{i.e}, 
$n-1$), the Boris scheme stores the velocity as a part of the usual macroparticle data structures. Thus, the modified Boris scheme is algebraically equivalent to these schemes.

We also note the work by Gordon et al.~\cite{Gordon2013}, which considers the coupling of linear dispersive media to PIC codes. In their approach, they advance the following equation:
\begin{gather}
\frac{\vec{x}_{b,n+1} - 2 \vec{x}_{b,n} + \vec{x}_{b,n-1}}{d t} \nonumber\\= \frac{q_a}{m_a} \vec{E}_n - \bm{\omega_b^2} \cdot \vec{x}_{b,n} - \bm{\gamma_b} \cdot \frac{\vec{x}_{a,n+1} - \vec{x}_{a,n}}{d t}.\label{eq:gordon}
\end{gather}
Here, particular attention should be given to the damping term, which is first-order accurate in contrast to the second-order accuracy used in the methods by Taflove, Varin, and our approach. While one might expect the first-order damping to introduce greater inaccuracies, our convergence tests—though not presented here—show that this method yields reasonable agreement with the expected results. However, it is slightly less accurate compared to the second-order accurate approaches, though it remains effective as long as the stability conditions discussed in \textcolor{blue}{Section 4.2} are adhered to.

We note that this implementation of the modified Boris scheme has advantages compared to the cell-based schemes in that this allows the re-use of the existing macroparticle concept and machinery already existing as a part of the particle pusher in these PIC codes. For example, if a problem requires higher order particle push, one need not re-implement higher order solvers which require extra subroutines for the separate resonator module but simply re-use the existing code that is used for the macroparticles. The same follows for current deposition and field interpolation routines. Finally, the fact that the linear dispersive response is implemented in macroparticles allows coupling between the bound particles and other physics available in the code. For example, a simulation can explore the use of collisions with bound particles, or ionization of bound particles to model laser damage of dispersive dielectrics. Therefore, the use of bound particles provides a number of remarkable capabilities for quantum plasma simulation in PIC.

Finally, as highlighted in previous works~\cite{Varin2015, Varin2018}, this model naturally accommodates the incorporation of nonlinear effects by adding higher-order terms in the electric field $\vec{E}_n$ to the Lorentz force, as described in Eq.~(\ref{eq:modlorentz}). This ability to capture nonlinearity is particularly valuable for modeling the behavior of nanophotonic structures. The extension of this model to include such nonlinear terms will be a key focus of future work.

\bibliographystyle{elsarticle-num}
\bibliography{maincpc.bib}

\end{document}